\documentclass[preprint]{revtex4-1}

\usepackage{amsmath}
\usepackage{amssymb}
\usepackage{amsfonts}
\usepackage{graphicx}
\usepackage{bbold}
\usepackage{microtype}

\usepackage[colorlinks=true,
            linkcolor=red,
            urlcolor=blue,
            citecolor=blue]{hyperref}

\begin{document}

\title{Faster than Classical Quantum Algorithm for dense Formulas of Exact
Satisfiability and Occupation Problems}

\author{Salvatore Mandr{\`a}}

\email{smandra@fas.harvard.edu}

\affiliation{Department of Chemistry and Chemical Biology, Harvard University,
12 Oxford Street, 02138 Cambridge (MA)}

\author{Gian Giacomo Guerreschi}

\email{guerreschi@fas.harvard.edu}

\affiliation{Department of Chemistry and Chemical Biology, Harvard University,
12 Oxford Street, 02138 Cambridge (MA)}

\author{Al{\'a}n Aspuru-Guzik}

\email{aspuru@chemistry.harvard.edu}

\affiliation{Department of Chemistry and Chemical Biology, Harvard University,
12 Oxford Street, 02138 Cambridge (MA)}
\begin{abstract}
We present an exact quantum algorithm for solving the Exact Satisfiability
problem, which belongs to the important NP-complete complexity class.
The algorithm is based on an intuitive approach that can be divided
into two parts: The first step consists in the identification and
efficient characterization of a restricted subspace that contains
all the valid assignments of the Exact Satisfiability; while the second
part performs a quantum search in such restricted subspace. The quantum
algorithm can be used either to find a valid assignment (or to certify
that no solution exists) or to count the total number of valid assignments.
The query complexities for the worst-case are respectively bounded
by $O\left(\sqrt{2^{n-M^{\prime}}}\right)$ and $O\left(2^{n-M^{\prime}}\right)$,
where $n$ is the number of variables and $M^{\prime}$ the number
of linearly independent clauses. Remarkably, the proposed quantum
algorithm results to be faster than any known exact classical algorithm
to solve dense formulas of Exact Satisfiability. As a concrete application,
we provide the worst-case complexity for the Hamiltonian cycle problem
obtained after mapping it to a suitable Occupation problem. Specifically,
we show that the time complexity for the proposed quantum algorithm
is bounded by $O\left(2^{n/4}\right)$ for 3-regular undirected
graphs, where $n$ is the number of nodes. The same worst-case complexity
holds for $(3,3)$-regular bipartite graphs. As a reference, the current
best classical algorithm has a (worst-case) running time bounded by
$O\left(2^{31n/96}\right)$. Finally, when compared to heuristic techniques
for Exact Satisfiability problems, the proposed quantum algorithm
is faster than the classical WalkSAT and Adiabatic Quantum Optimization
for random instances with a density of constraints close to the satisfiability
threshold, the regime in which instances are typically the hardest
to solve. The proposed quantum algorithm can be straightforwardly
extended to the generalized version of the Exact Satisfiability known
as Occupation problem. The general version of the algorithm is presented
and analyzed.
\end{abstract}
\maketitle

\section{Introduction}

Constraint satisfaction problems (CSPs) play a fundamental role in
both theoretical and applied computer science. Even though the specific
formulation of such problems may vary, all CSPs are characterized
by a certain number $M$ of clauses (or constraints) involving $n$
boolean variables: An assignment is valid if and only if all the clauses
are satisfied. The central question of many CSPs is either to exhibit
a valid assignment or to prove that there is none. Despite its simplicity,
answering this question is a hard task. Accordingly, many CSPs belong
to the class of NP-complete problems \cite{karp1972reducibility,schaefer1978complexity},
namely the class of the hardest decision problems whose solution can
be efficiently verified. The first problem proved to belong to the
NP-complete class was the Satisfiability (SAT) problem \cite{cook1971complexity}.
In SAT problems, clauses are composed by variables which may appear
either negated or not, and each clause is satisfied if it contains
at least one $true$ literal. The problem remains NP-complete even
if any clause contains only three variables (3SAT) \cite{karp1972reducibility}.
Although no theorem has yet confirmed this assumption, it is widely
believed that no polynomial time algorithm exists to solve NP-complete
problems: This is the famous question of P being different from NP.
Indeed, the computational time for either heuristics (namely those
algorithms which may provide a valid assignment in case of success,
but without certifying that there is none in case of their failure)
or exact algorithms (which always provide such certification) is expected
to scale exponentially with the system size $n$.

Quantum computing can be used to solve classical problems and, in
particular, CSPs. In general, quantum algorithms are believed to have
a better scaling than their classical counterparts. A few remarkable
examples, like integer factorization \cite{shor1997polynomial}, pattern
matching \cite{montanaro2014quantum} or solving systems of linear
equations \cite{harrow2009quantum,ambainis2012variable}, achieve
an exponential speedup over the known best classical alternative.
However, it seems that only a polynomial speedup can be reached by
quantum algorithms in the context of NP-complete problems, meaning
that NP-complete problems remain hard to solve for both classical
and quantum computers. 

The Occupation problem (also called $q$-in-$p$-SAT) is a variant
of the general Satisfiability (SAT) problem \cite{garey1979}. In
this case, clauses are satisfied if and only if they contain exactly
$q\ge1$ $true$ literal. The case of $q=1$ is also known as Exact
Satisfiability (XSAT) and has been extensively studied \cite{schaefer1978complexity}.
Both Occupation and XSAT problems belong to the NP-Complete complexity
class, even in the restricted case in which all the literals occur
only unnegated \cite{schaefer1978complexity}. Recently, 1in3-SAT
(also known as X3SAT or Exact Cover) and 2in4-SAT problems have been
used for the benchmark of both classical and quantum optimization
algorithms \cite{young2008size,farhi2001quantum,hen2011exponential,guidetti2011complexity}
and for the theoretical understanding of the satisfiability transition
in correspondence to the critical density of constraints $\alpha_{SAT}$
\cite{zdeborova2008constraint} (specifically, $\alpha_{SAT}$ is
the maximum $\alpha=M/n$ such that randomly chosen instances have,
with high probability, a solution in the limit of large number of
variables). Indeed, the definition of NP-completeness regards the
worst-case scenario while it has been shown that the randomly chosen
instances are typically hard only close to the satisfiability transition
$\alpha_{SAT}$ \cite{cheeseman1991really,mezard2002analytic,krzakala2007gibbs}.

\begin{figure}
\begin{centering}
\includegraphics[width=0.45\textwidth]{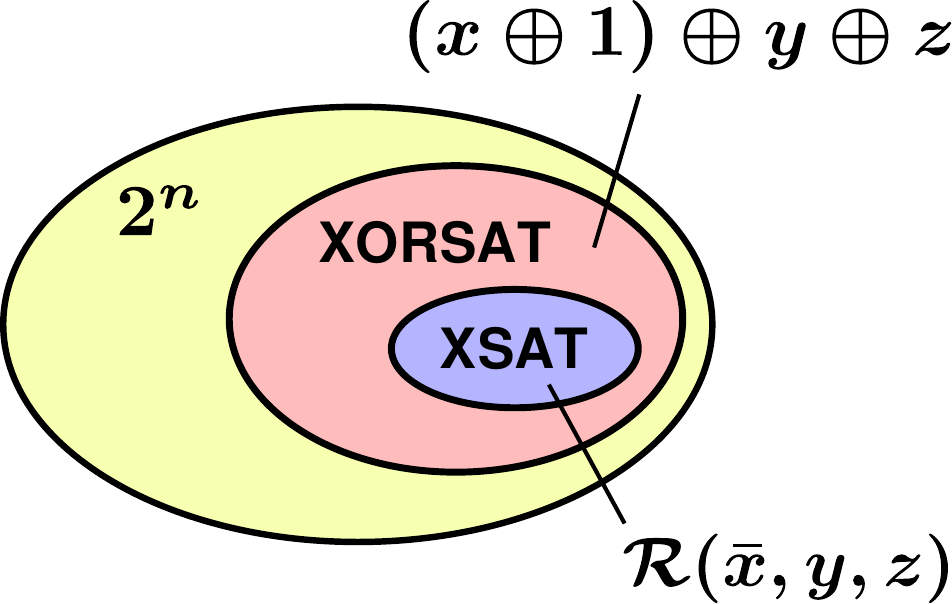}
\par\end{centering}

\caption{\textbf{\label{fig:XSAT_reduction}A necessary condition for a given
assignment to satisfy a generic XSAT or Occupation problem clause
is that the same assignment satisfies the corresponding XORSAT clause.}
Here, for example, $\mathcal{\mathcal{R}}(\bar{x},\,y,\,z)$ represents
a 1in3-SAT clause that is $true$ if and only if exactly one between
$\bar{x}$, $y$ and $z$ is $true$. As a consequence, inside the
set of all the possible $2^{n}$ assignments for the $n$ variables,
there is an intermediate set that contains all the solutions of a
specific XSAT instance and that represents all the solutions to a
related XORSAT instance.}
\end{figure}
In the last three decades, the upper bound for the computational time
of exact algorithms for both XSAT and 1in3-SAT have progressively
improved. The first algorithm proved to be faster than the trivial
$2^{n}$ scaling (corresponding to the exhaustive enumeration of all
the possible assignments) was provided by Schroeppel and Shamir \cite{schroeppel1981t}.
Their algorithm solves a class of problems, of which XSAT is the most
relevant, in time $O(2^{n/2})$ but in exponential space $O(2^{n/4})$.
In the same year, a better result was provided by Monien, Speckenmeyer
and Vornberger with an algorithm which runs in time $O(2^{0.2441n})$
and requires polynomial space \cite{monien1981upper}. Later on, the
upper bound for the computational complexity for both XSAT and 1in3-SAT
has further lowered \cite{porschen2005exact,kulikov2002upper}. Currently,
the best exact algorithm has been proposed by Byskov, Madsen and Skjerna,
which solves XSAT and 1in3-SAT respectively in time $O(2^{0.2325n})$
and $O(2^{0.1379n})$ \cite{byskov2005new} and polynomial space.
The better scaling has been achieved by branching only on those variables
which appears in at least three clauses and solving the remaining
problem as a perfect matching problem. Since the perfect matching
is in the P-Class, what remains after the branching can then be solved
in polynomial time. Surprisingly, there are only few algorithms for
which the complexity is computed in terms of the number of clauses
$M$. Among them, Skjernaa presented an algorithm for XSAT with a
time bound $O(2^{M})$, but using exponential space \cite{skjernaa2004exact}.
A better algorithm which uses polynomial space and time $O(2^{0.2123M})$
has been provided by Zhou and Yin \cite{zhou2012worst}. Excluding
the algorithm by Schroeppel and Shamir \cite{schroeppel1981t}, all
the algorithms mentioned above are \emph{branch-and-reduce} algorithms,
which means that the variables (or alternatively the clauses) are
iteratively removed before a reduction step \cite{davis1960computing,davis1962machine}.

While algorithms for many NP-Complete problems are well studied little
is known for the case of their \#P-Complete counterparts \cite{valiant1979complexity},
namely the problem of counting the number of valid assignments. The
counting version of many satisfiability problems is not only interesting
from the mathematical point of view, but it has important applications
in other fields like artificial intelligence \cite{birnbaum1999good,roth1996hardness}
and simulation of physical systems (see, for example, the recent problem
of boson sampling \cite{aaronson2011computational,tillmann2015,huh2014boson}).
At the moment, the fastest algorithm for \#XSAT and \#1in3-SAT are
respectively bounded by $O(2^{0.262n})$ \cite{zhou2014new} and $O(2^{0.197n})$
\cite{dahllof2004algorithms}.

In this work, we present an exact algorithm for the XSAT and Occupation
problems which runs, in the worst-case, in a time bounded by $O(2^{n-M^{\prime}})$
on classical computer and by $O(\sqrt{2^{n-M^{\prime}}})$ on a quantum
computer, while using only polynomial space. Here, $M^{\prime}\leq\min\left\{ n,\,M\right\} $
is the number of linearly independent clauses. In the quantum case,
with the term \emph{exact} we do not intend that the algorithm is
\emph{deterministic,} but rather that the quantum algorithm has a
bounded error to either provide a solution or certify that none exists
which decreases exponentially with the number of repetitions.

For the corresponding counting problem, our classical and quantum
algorithms are, respectively, bounded by $O(2^{n-M^{\prime}})$ and
$O(\sqrt{V\,2^{n-M^{\prime}}})$, where $V\leq2^{n-M^{\prime}}$ is
the number of valid assignments not known \emph{a priori}. We arrived
to this result by introducing a novel reduction of the Occupation
problem in order to identify a subset of assignments that is guaranteed
to contain all the solutions (see Figure~\ref{fig:XSAT_reduction}).
Such reduction, that we refer to with the name XOR-Reduction, differs
from the branch-and-reduce approach used in the past and leads to
an efficient way to find and characterize the appropriate subset.
The quadratic speedup for the quantum version of the algorithm is
achieved by performing a Grover-type search \cite{grover1996fast,farhi1998analog,roland2002quantum,mandra2015adiabatic}
among the candidate solutions. As the main result, we found that our
quantum algorithm is faster than the (known) best exact classical
algorithms for dense formulas. For specific problems, it is even possible
to find a lower bound for $M^{\prime}$ expressed as a function of
$n$ only. As a concrete example, we consider the Hamiltonian cycle
problem, namely the problem to find a closed and non-intersecting
path that explores all the nodes of a given graph. Specifically, we
show that the worst-case complexity for 3-regular undirected graphs
is bounded by $O\left(2^{n/4}\right)$, where $n$ is the number
of nodes of the graph. The same bound holds for $(3,3)$-regular bipartite
graphs. As a comparison, the (known) best classical algorithm is bounded
by $O\left(2^{31n/96}\right)$ \cite{IwamaN07}.

In addition, we analyze the performance of the proposed algorithm
on random instances of Occupation problems to evaluate its computational
cost for the typical-case. This analysis is particularly significant
to estimate the actual advantage of our quantum algorithm in real
world situations which, arguably, is represented by typical, rather
than worst-case, instances. We observe that, differently from the
worst-case complexity, the typical computational cost is reduced when
additional strategies inspired by the classical backtracking technique
are applied \cite{van2006backtracking,montanaro2015quantum}. Thus,
we show that the proposed quantum algorithm still remains the fastest
one (compared to classical heuristics like WalkSAT or quantum heuristics
like Adiabatic Quantum Optimization) for random instances of 1in3-SAT
and 2in4-SAT problems close to the satisfiability threshold \cite{hen2011exponential,guidetti2011complexity},
where the instances are typically the hardest ones to solve.

The paper is organized as follows: In Section~II we present and characterize
our XOR-Reduction algorithm. Section~III is dedicated to the analysis
of the worst-case complexity and includes the explicit construction
of the quantum oracle involved in the Grover search. The explicit
application of our algorithm to the Hamiltonian cycle problem is the
topic of Section~IV, while in Section~V we discuss classical and
quantum backtracking techniques and evaluate the typical computational
cost of the proposed algorithm. Finally, in the last Section we provide
additional discussions and draw conclusions.

\section{Reduction Method for XSAT and Occupation Problems}

\label{sec:reduction-method}Any Occupation ($q$-in-$p$-SAT) problem
instance is composed of $n$ variables and $M$ constraints, whereas
each constraint accepts exactly $p$ variables as input. More specifically,
let $\mathcal{R}(x_{1},\,x_{2},\,\ldots,\,x_{p})$ be a boolean function
that is $true$ if and only if exactly $q$ among all the $p$ literals
$x_{i}$ are $true$. An arbitrary instance $\mathcal{I}$ of the
Occupation problem can be written as 
\begin{equation}
\mathcal{I}\left(x_{1},\,\ldots,\,x_{n}\right)=\bigwedge_{a=1}^{M}\mathcal{R}\left(\tilde{x}_{a_{1}},\,\tilde{x}_{a_{2}},\,\ldots,\,\tilde{x}_{a_{p}}\right),\label{eq:def_X3SAT}
\end{equation}
where $\{a_{1},\,a_{2},\,\ldots,\,a_{p}\}$ indicates what variables
are involved in the constraint $a$, and the symbol $\sim$ means
that the variable may appear either negated or unnegated. In the rest
of the paper, we will focus on \emph{locked} instances of the Occupation
problem, where all the variables appear in at least two clauses. Indeed,
variables which appear in only one clause can be iteratively removed
from the problem, while locked instances are typically hard to solve
\cite{zdeborova2008constraint,zdeborova2008locked}.

At the core of both the classical and quantum algorithms that we propose
lies the observation that $\mathcal{R}$ is satisfied only if the
corresponding XORSAT clause is also satisfied:
\begin{equation}
\mathcal{R}_{XOR}\left(\tilde{x}_{a_{1}},\,\tilde{x}_{a_{2}},\,\ldots,\,\tilde{x}_{a_{p}}\right)=\tilde{x}_{a_{1}}\oplus\tilde{x}_{a_{2}}\oplus\ldots\oplus\tilde{x}_{a_{p}}\oplus r,\label{eq:X3SAT_reduction}
\end{equation}
where $\oplus$ represents the XOR sum (namely, the sum of integer
modulo 2 using the convention $0\equiv false$ and $1\equiv true$)
and $r=\left(q+1\right)_{\text{mod}\,2}$. Therefore, any valid assignment
$\left\{ x_{1},\,\ldots,\,x_{n}\right\} $ for $\mathcal{I}$ must
be a valid assignment for the corresponding XORSAT instances, namely:
\begin{align}
\mathcal{I}_{XOR}\left(x_{1},\,\ldots,\,x_{n}\right)= & \bigwedge_{a=1}^{M}\left(\tilde{x}_{a_{1}}\oplus\tilde{x}_{a_{2}}\oplus\ldots\oplus\tilde{x}_{a_{p}}\oplus r\right).\label{eq:3XOR}
\end{align}
It is important to observe that, unlike the original $q$-in-$p$-SAT
problem, there exists a classical algorithm which solves the XORSAT
problem in polynomial time. Indeed, any valid assignment for $\mathcal{I}_{XOR}$
is also a solution of the following non-homogeneous linear problem
\cite{lidl1997finite,kolchin1999random}:
\begin{equation}
\left(Ax\equiv b\right)_{\textrm{mod}\,2},\label{eq:lin_eq}
\end{equation}
where $A$ is a $M\times n$ binary-matrix, for which $A_{ai}=1$
if and only if variable $i$ appears (either negated or not) in clause $a$,
and is zero otherwise. Here, $b_{a}=\left(\nu_{a}+q\right)_{\textrm{mod}\,2}$
with $\nu_{a}$ being the number of negations that appear in the clause
$a$. To make the description more concrete, for the locked 1in3-SAT
class of instances, the matrix $A$ results to be sparse with exactly
three ones in each row and at least two ones in each column. 

\begin{figure}
\begin{centering}
\includegraphics[width=0.65\textwidth]{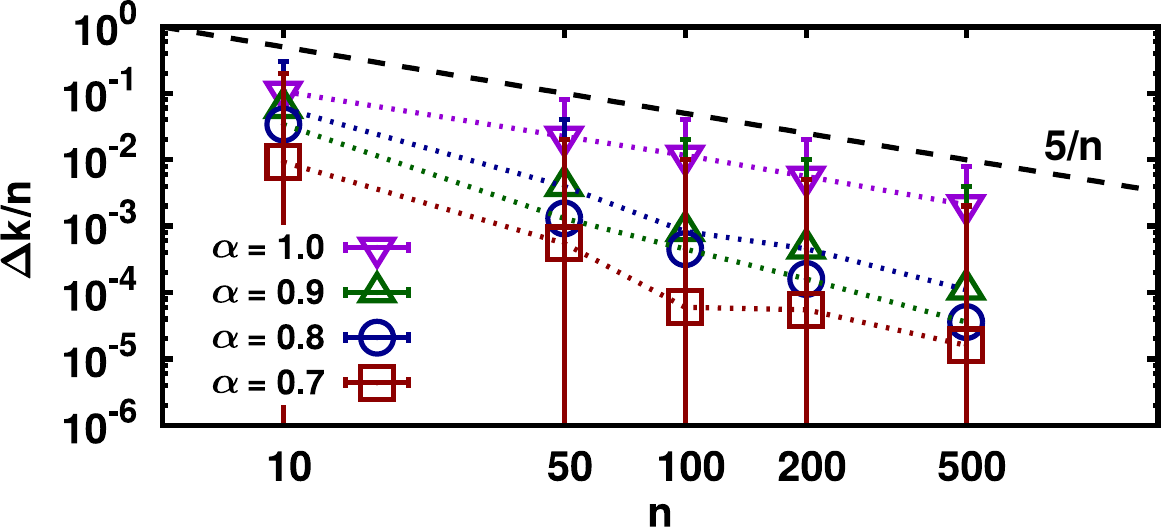}
\par\end{centering}

\caption{\textbf{\label{fig:scaling_kernel}The number $\boldsymbol{k}$ of
linearly independent vectors of the kernel quickly converges to $(\boldsymbol{n-M})$,
where $\boldsymbol{n}$ and $\boldsymbol{M}$ are respectively the
number of variables and constraints.} The figure shows the excess
of the number of linearly independent vectors for locked 1in3-SAT
instances, namely $\Delta k=k-(n-M)$. As one can see, $\Delta k/n$
goes to zero by increasing the number of variables. The analysis is
based on $1000$ random realization of locked 1in3-SAT instances for
each $n$ and $\alpha$. Points and error bars represent respectively
the average and the minimum/maximum of the numerical distribution
of $\Delta k/n$. Notice that, as clearly indicated by the comparison
with the dashed black line, we obtain $\Delta k\leq5$ independently
of $n$ and even for $\alpha=1$.}
\end{figure}
Let us define $\left\{ \xi_{1},\,\xi_{2},\,\ldots,\,\xi_{k}\right\} $
as the set of $k-$independent vectors which solve the following kernel
equation
\begin{equation}
\left(A\xi_{k}\equiv0\right)_{\textrm{mod}\,2},\label{eq:kern_lin_eq}
\end{equation}
and denote by $M^{\prime}=(n-k)\leq\min\left\{ n,\,M\right\} $ the
number of linearly independent (with arithmetic modulo 2) rows of
the matrix $A$. Therefore, $M^{\prime}$ represents the number of
independent clauses. It is important to stress that all the theorems
of linear algebra are still valid in arithmetic modulo 2. Hence, the
set $\left\{ \xi_{1},\,\xi_{2},\,\ldots,\,\xi_{k}\right\} $ can be
found in polynomial time using a classical computer (for example,
via Gaussian elimination \cite{lidl1997finite}). Moreover, all the
solutions of Equation~(\ref{eq:lin_eq}), i.e. all the valid assignment
for $\mathcal{I}_{XOR}$, can be expressed as
\begin{equation}
x\left(v_{1},\,\ldots,\,v_{k}\right)=v_{1}\xi_{1}\oplus v_{2}\xi_{2}\oplus\ldots\oplus v_{k}\xi_{k}\oplus\bar{\xi},\label{eq:XOR_space}
\end{equation}
where $v_{i}\in\left\{ 0,\,1\right\} $ and $\bar{\xi}$ is a particular
solution (of the inhomogeneous part) of Equation~(\ref{eq:lin_eq}),
that is to say $A\bar{\xi}=b$. This implies that the number of valid
assignment for $\mathcal{I}_{XOR}$ is $2^{k}=2^{n-M^{\prime}}$.
Observe that, given the \emph{rank-nullity} theorem, the number of
independent rows $M^{\prime}$ is always smaller or equal to the total
number of clauses $M$. Moreover, if $A$ is a sparse random matrix,
it happens that $M^{\prime}/M=O(1)$ in the limit of large $n$ and
fixed $\alpha=M/n$ \cite{kolchin1999random,alamino2008typical,alamino2009properties}
(see Figure~\ref{fig:scaling_kernel} for a numerical confirmation
for locked instances of the 1in3-SAT problem).

\section{Worst-case Analysis for the Proposed Classical and Quantum Algorithms}

\label{sec:worst-case}The classical algorithm that we introduce with
this work relies on the fact that the configuration space of all the
valid assignments for $\mathcal{I}_{XOR}$, which is given by Equation~(\ref{eq:XOR_space}),
is actually much smaller than the whole configuration space associated
with $n$ variables. Consequently, it is possible to find a valid
assignment for any specific instance $\mathcal{I}$ of the Occupation
problem by (i) enumerating all the possible valid assignments for
$\mathcal{I}_{XOR}$ and (ii) checking which of these is also a valid
assignment for $\mathcal{I}$. Since $\bar{\xi}$ and all the $\xi_{i}$
can be find in polynomial time and stored in polynomial space, the
classical algorithm uses polynomial space and its run-time is determined
by the cost of the search. Even in the worst-case scenario of an exhaustive
search, the number of operations is bound by the dimension of the
configuration space of the valid assignments for $\mathcal{I}_{XOR}$.
Therefore, our algorithm requires at most $O(2^{n-M^{\prime}})$ calls
of the oracle that distinguishes invalid assignments from actual solutions,
where we use the previous definition of $M^{\prime}\leq\min\left\{ n,\,M\right\} $
as the number of independent clauses for the XORSAT problem. Finally,
it is straightforward to verify that the proposed algorithm is \emph{exact,}
since a valid assignment for $\mathcal{I}$ must also be a valid assignment
for $\mathcal{I}_{XOR}$, and that the complexity for the corresponding
counting problem remains $O(2^{n-M^{\prime}})$.

\begin{figure}
\begin{centering}
\includegraphics[width=0.95\textwidth]{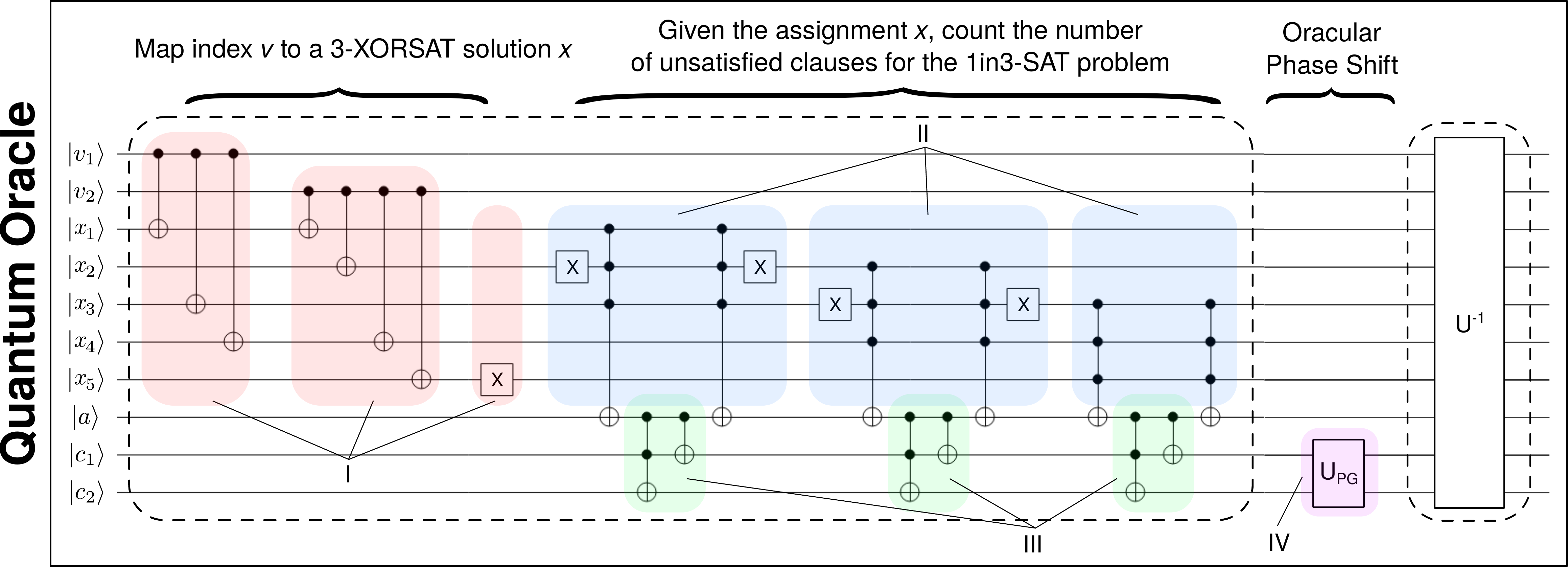}
\par\end{centering}

\caption{\textbf{\label{fig:example_qcirc}Example of our quantum algorithm
applied to solve an instance of the 1in3-SAT problem} $\mathbf{\mathcal{R}(x_{1},\bar{x}_{2},\,x_{3})\wedge\mathcal{R}(x_{2},\bar{x}_{3},\,x_{4})\wedge\mathcal{R}(x_{3},x_{4},\,x_{5})}$,
where $\mathcal{\mathcal{R}}(a,\,b,\,c)$ is $true$ if and only if
exactly one between $a$, $b$ and $c$ is true. A detailed description
of the quantum algorithm is reported in the main text.}
\end{figure}
The proposed exact quantum algorithm for the Occupation problem is
obtained by substituting the exhaustive check of all valid assignments
of $\mathcal{I}_{XOR}$ with a Grover-type quantum search \cite{grover1996fast}.
This search actually takes place in the smaller register of $k$ qubits
in which all the candidate solutions of $\mathcal{I}$ are identified
by the corresponding $k$ bits $\left\{ v_{1},\,\ldots,\,v_{k}\right\} $.
Due to this abstract representation, the initial state preparation
(superposition of all valid assignments of $\mathcal{I}_{XOR}$) and
the Grover diffusion operator coincide with the original prescription
of the Grover algorithm and are easily implemented, for example, with
$O\left(k\right)$ Hadamard gates and a single $k$-qubit phase gate.
Finally, one has to implement an oracle for $\mathcal{I}$ that requires
only a polynomial number of operations, which is always possible since
the Occupation problem belongs to the NP class (we provide an explicit
construction in Figure~\ref{fig:example_qcirc} and analyze the number
of gates required at the end of this Section). More specifically,
the quantum oracle $\hat{\mathcal{O}}$ corresponds to the operator
defined by
\begin{equation}
\hat{\mathcal{O}}\left|v\right\rangle =\begin{cases}
-\left|v\right\rangle  & \text{if }\mathcal{I}\left(x(v)\right)\textrm{ is true}\\
+\left|v\right\rangle  & \textrm{otherwise}
\end{cases},\label{eq:quantum_oracle}
\end{equation}
with $x$ given by Equation~(\ref{eq:XOR_space}). The oracle $\hat{\mathcal{O}}$
is central to the application of the Grover algorithm that finds a
valid assignment for $\mathcal{I}$ and, as we show below, it can
be implemented with only a polynomial overhead. Therefore, the Grover
algorithm results to be quadratically faster than a classical exhaustive
search in the reduced XORSAT solution space \cite{grover1996fast}.
The same speedup holds even in the presence of multiple valid assignments
\cite{boyer1996,brassard1998quantum,brassard2002quantum,hayashi2014introduction}
and represents the optimal speedup achievable for unstructured searches.
Therefore, the prosed quantum algorithm uses only polynomial space
and it is bounded in time by $\mathcal{O}(\sqrt{2^{n-M^{\prime}}})$
gate operations. It is important to stress that the proposed quantum
algorithm is \emph{exact}, namely it is possible to certify that no
valid assignments exist: This property is based on the possibility
of running a generalized Grover quantum algorithm that preserves the
quadratic speedup even if the number of target states (i.e. our valid
assignments) is not known a priori, but larger then zero \cite{brassard2002quantum}.
The case in which no solution is actually present can be included
by slightly modifying the quantum oracle $\hat{\mathcal{O}}$ and
forcing it to accept a specific configuration as a valid assignment
and verifying that this represents the only (in this case ``artificial'')
solution.

\begin{figure}
\begin{centering}
\includegraphics[width=0.75\textwidth]{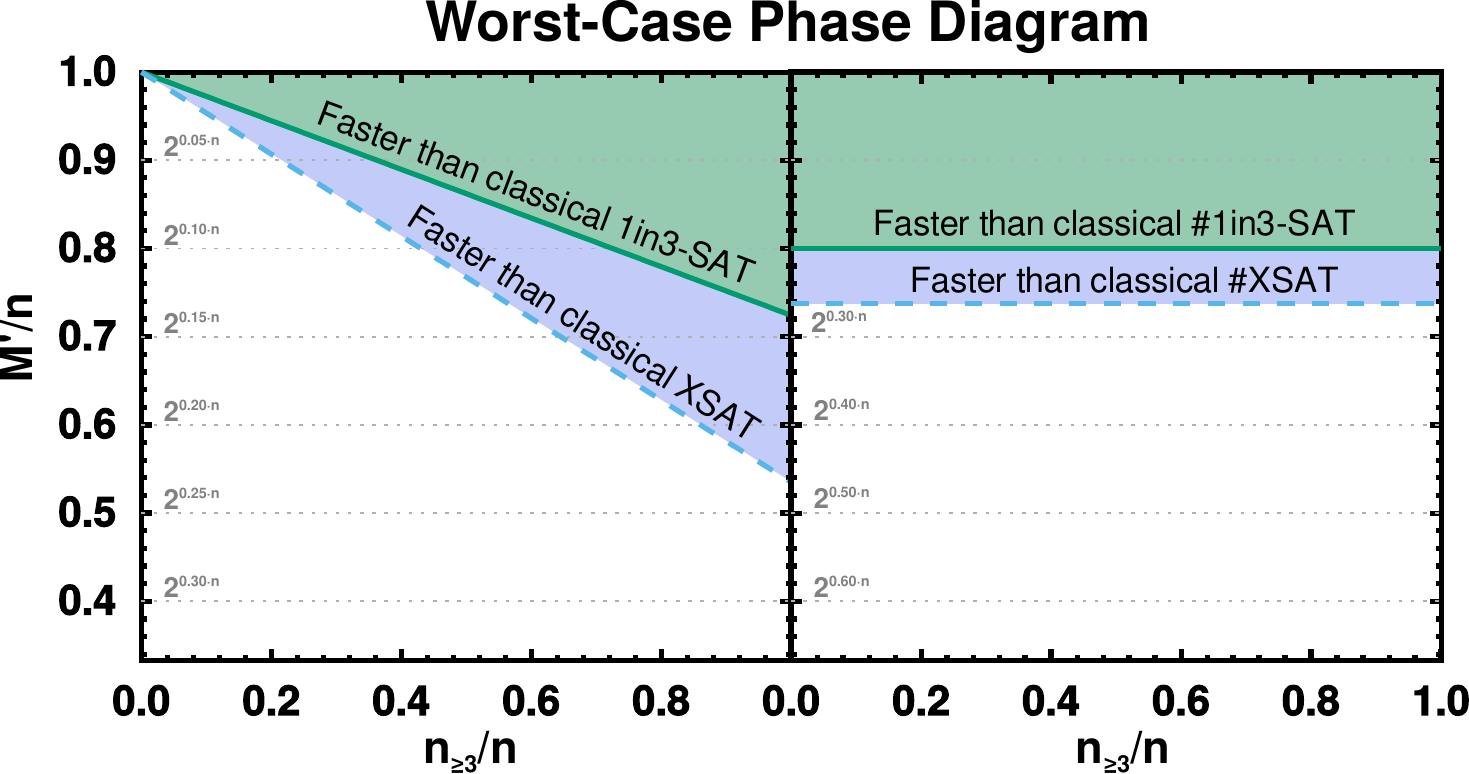}
\par\end{centering}

\caption{\textbf{\label{fig:scaling_worst_case}The proposed quantum XOR-Reduction
algorithm is faster than the known best exact classical algorithms
for dense formulas.} The figure shows the phase diagram (based on
the worst-case scenario) that identifies the regions in which the
proposed quantum algorithm outperforms the best classical algorithms
known to date, either to solve XSAT and 1in3-SAT instances (left panel)
or to count the number of valid assignments (right panel). Here and
in the main text, $n$ is the number of variables, $n_{\geq3}$ is
the number of variables which enter in at least three clauses and
$M^{\prime}$ is the number of independent clauses (as defined in
the main text). In gray, the complexity for the proposed quantum algorithm
(represented by horizontal lines since it does not depend on $n_{\geq3}$). }
\end{figure}
The Grover algorithm, with the same quantum oracle $\hat{\mathcal{O}}$,
can also be applied for counting the number of valid assignments \cite{brassard1998quantum,brassard2002quantum}.
In this case, the number of calls of $\hat{\mathcal{O}}$ is bounded
by $O(\sqrt{V\,2^{n-M^{\prime}}})$, where $V\leq2^{n-M^{\prime}}$
is the number of solutions of the Occupation problem. Unfortunately,
in the trivial but worst-case scenario when all assignments are solutions,
the complexity for the \#P problem of the proposed quantum algorithm
is bounded by $O(2^{n-M^{\prime}})$, as for its classical counterpart.

In order to provide an explicit example, in Figure~\ref{fig:scaling_worst_case}
we compare the complexity of the proposed quantum to the (known) best
classical algorithms for XSAT/1in3-SAT \cite{byskov2005new} (left
panel) and \#XSAT/\#1in3-SAT \cite{zhou2014new,dahllof2004algorithms}
(right panel). In the comparison, we also take into account that the
complexity for the best classical XSAT/1in3-SAT algorithm depends
only in the number of variables which enter in at least three clauses
$n_{\geq3}$, namely $O(2^{0.1379\,n_{\geq3}})$. 
As it has been originally shown by Cheeseman \emph{et al.} \cite{cheeseman1991really}
for the SAT problem (and successively for many other satisfiability
problems including the Exact Satisfiability problem \cite{zdeborova2008constraint}), 
random instances are actually hard
only when the problems' parameters are chosen so that 
instances have a similar probability to either have a valid assignment or not.
In most cases, the relevant parameter is the density of constraints
$\alpha = M/n$: Indeed, for small $\alpha$, random instances
are likely to have a valid assignment while, for large $\alpha$, random instances
most likely do not have any valid assignment. On the contrary, at the satisfiability
threshold $\alpha_{SAT}$, random instances might have or not a valid assignment
with comparable probability. Interestingly,
our quantum algorithm has a better scaling for dense formulas characterized
by a sufficiently large density of independent constraints $M^{\prime}/n$,
far away from the simple case of low $\alpha$.
Our algorithm results particularly fast for those instances
which are, instead, expected to be hard for the classical XSAT/1in3-SAT
algorithms (see Section~\ref{sec:typical} for the analysis
of the proposed quantum algorithm by using random instances extracted
at the satisfiability threshold).
Also the vice-versa seems true, as indicated by the apparent
complexity of the trivial case in which the density of constraints
is small. We discuss this fact more extensively in the final discussions.
A careful reader could observe that the computational complexity of
the (known) best classical algorithm for the XSAT/1in3-SAT is expressed
only in terms of number of variables $n$ and then, it might have
a better computational complexity if the number of linearly independent
clauses $M^{\prime}$ would be taken into account. However, the classical
algorithm is expected to perform the worse when the density $n_{\geq3}/n$
becomes large, which is exactly the region where the proposed quantum
algorithm performs the best.

To conclude this Section, we provide an explicit construction for
the quantum oracle $\hat{\mathcal{O}}$ that is polynomial in both
space (i.e. number of ancilla qubits) and time (i.e. number of gates).
The quantum oracle accepts $k$ qubits as input $\left\{ \left|v_{1}\right\rangle ,\,\ldots,\,\left|v_{k}\right\rangle \right\} $
(see Figure~\ref{fig:example_qcirc} for a graphical representation),
so that a computational basis state represents a specific valid assignment
for $\mathcal{I}_{XOR}$ as described in Equation~(\ref{eq:XOR_space})).
The oracle $\hat{\mathcal{O}}$ \emph{per se} is composed by four
modules, as depicted in Figure~\ref{fig:example_qcirc}. Their action
is best described in the computational basis, even if they can (and
obviously will) be applied to an initial superposition: (I) A module
to construct the $n$-qubit state $\left|x\right\rangle $ as given
in Equation~(\ref{eq:XOR_space}) from the corresponding valid assignments
for $\mathcal{I}_{XOR}$, (II) a module to verify if clauses of $\mathcal{I}$
are satisfied, (III) a repeated module to count the number of unsatisfied
clauses for $\mathcal{I}$ and, finally, (IV) a multi-qubit phase
gate that add a multiplicative phase $(-1)$ if and only if the number
of satisfied clauses for $\mathcal{I}$ is $M$. All the modules require
only the use of C-NOT gates and $X$ gates. The suggested implementation
of the quantum oracle requires $\left\lceil \log\,M^{\prime}\right\rceil $
ancilla qubits, while the total number of gates is bounded by $O\left(n^{2}\right)$
since (I) has $k$ blocks with $O\left(n\right)$ gates each, (II)
has $M^{\prime}$ blocks with $O\left(1\right)$ gates each and (III)
has $M^{\prime}$ blocks with $O\left(\log_2 M^\prime\right)$ gates each.

\section{Application to the Hamiltonian Cycle Problem}

The Hamiltonian cycle (HC) problem, together with the SAT problem,
can be considered one of the fundamental and most studied NP-complete
problem \cite{garey1979computers}. The HC problem consists in finding
a closed path (namely, an Hamiltonian cycle) which ``visits'' once,
and only once, all the nodes of a given $n$-nodes graph $\mathcal{G}$.
Unlike the SAT problem, which is called a \emph{subset} problem whose
configuration space is $2^{n}$, HC is a \emph{permutation} problem
whose configuration space is $n!$ \cite{IwamaN07}. Compared to SAT,
HC results much harder to solve and, at the moment, no algorithm has
been found to obtain Hamiltonian cycles for arbitrary graphs with
worst-case complexity better than $O\left(2^{n}\right)$ \cite{held1962dynamic}.
However, improved scalings have been obtained for bounded degree graphs
\cite{eppstein2007traveling,IwamaN07}. For example, for 3-regular
graphs, the (known) best classical algorithm has the worst-case complexity
bounded by $O\left(2^{31n/96}\right)$ \cite{IwamaN07}. In this Section
we show that our quantum algorithm can find a Hamiltonian cycle in
a time bounded by $O\left(2^{(K-2)\,n/4}\right)$, where $K$
is the maximum degree of $\mathcal{G}$. This result can be achieved
by reducing the HC problem to an Occupation problem. In case of bounded
graphs with $K = 3$, the worst-case complexity is given by $O\left(2^{n/4}\right).$
The same worst-case complexity holds for $(3,3)$-regular bipartite
graphs. For the rest of the Section, it is assumed that the graph
$\mathcal{G}$ is an undirected graph with a single connected component
and all the nodes having the same degree equals to $K$ (to avoid
trivial cases, we consider $K\geq3$). Results for bipartite graphs
follow directly.

As a first step to apply the proposed XOR-Reduction, it is necessary
to reduce the HC problem to an Occupation problem. The simplest way
to achieve this is to define a variable $e_{ij}$ for any of the $M$
edges of $\mathcal{G}$ such that $e_{ij}=e_{ji}=1$ if and only if a given
path passes through the edge which connects node $i$ and $j$, and
zero otherwise. Therefore, any possible path in $\mathcal{G}$ can
be expressed as a specific assignment of $\left\{ e_{ij}\right\} =\left\{ 0,\,1\right\} ^{M}$.
Recalling that an Hamiltonian cycle must visit once and only once
each node of $\mathcal{G}$, the given path satisfies the HC problem
if it also satisfies the set of extra constraints
\begin{equation}
\sum_{j\in\partial i}e_{ij}=2,\ \forall i\in\mathcal{G},\label{eq:HC_xsat}
\end{equation}
where $\partial i$ represents the set of nodes connected to $i$.
Clearly, the constraints in Eq.~(\ref{eq:HC_xsat}) represent 2in$K$-SAT
clauses, one for each node. Therefore, given a graph $\mathcal{G}$,
all valid Hamiltonian cycles must be the solution of a 2in$K$-SAT
instance with $M$ variables and $n$ clauses (please, notice the
different notation in which $n$ refers to the number of clauses and
not, as in all other Sections, of variables). Also, observe that the
related 2in$K$-SAT instance may have solutions which do not correspond
to any Hamiltonian cycle (for example, a collection of separate cycles
rather than a single long cycle). However, it suffices to change the
quantum oracle $\hat{\mathcal{O}}$ so that it can check if a solution
of the 2in$K$-SAT instance is also a Hamiltonian cycle, with only
a polynomial overhead. Hence, following the results of Section~\ref{sec:worst-case},
the proposed quantum algorithm can either find an Hamiltonian cycle
or certify that none exists in a time bounded by $O\left(\sqrt{2^{M-n^{\prime}}}\right)=O\left(\sqrt{2^{\frac{K}{2}n-n^{\prime}}}\right)$,
where $n^{\prime}$ is the number of ``linearly independent'' nodes.
The rest of the Section is dedicated to the proof that $n^{\prime}=n-1$.\\

In general, a set of vectors is linearly dependent if their sum (modulo
2) gives a vector of all zeros. However, if a vector of the set has
at least one unmatched component set to 1, namely none of other vectors
has a 1 in the same position, such vector is trivially linearly independent
from the others. Therefore, in the search for linearly dependent vectors,
it can be ``removed'' from the set since none of the other vectors
can linearly depend on it. For a given a graph $\mathcal{G}$, the
corresponding matrix $A$ of the 2in$K$-SAT instance corresponds
to a $n\times M$ matrix with exactly $K$ ones each row (since exactly
$K$ edges are connected to each node) and $2$ ones each column (every
edge connects two nodes). In order to prove that $n^{\prime}=n-1$,
we will show that all the rows of $A$ except one are trivially linearly
independent. Indeed, given that all columns contain exactly two ones,
the matrix $A$ has at least one linearly dependent row (this can
be verified since the sum modulo 2 of all rows gives the null vector).
Once such linearly dependent row is removed from $A$, it is straightforward
to see that $K$ rows will have an unmatched ones that can be removed
since trivially linearly independent. The iterative process to remove
rows from $A$ with unmatched ones proceeds until all the rows associated
to nodes in a connected component are removed. Hence, recalling that
$\mathcal{G}$ has single connected component, the process stops only
after the matrix $A$ is empty. Consequently, all the rows of $A$
must be linearly independent expect for the first one, namely $n^{\prime}=n-1$.

\section{Typical Case Analysis for the Quantum Algorithm}\label{sec:typical}

In Section~\ref{sec:worst-case}, we have provided upper bounds for
the worst-case complexity of the proposed quantum XOR-Reduction algorithm
in two situations: Either as a decision problem in which the algorithm
exhibits a valid assignment (or certifies that none exists), or as
a counting problem in which the algorithm provides the number of solutions.
More precisely, we used the Grover algorithm to show that the number
of query calls are bounded by $O\left(\sqrt{2^{n-M^{\prime}}}\right)$
and by $O\left(2^{n-M^{\prime}}\right)$ respectively, where $n$
is the number of variables and $M^{\prime}$ the number of independent
clauses. It is important to observe that the Grover algorithm is completely
insensitive to any structure of the underlying problem. In many cases,
there exist correlations between variables and clauses of a specific
instance which can effectively reduce the size of the configuration
space to explore. Classical algorithms are constructed to take advantage
of such sort of information with the result that their typical performance
is better than what the worst-case upper bound would suggest.

One of the most important and successful classical techniques to solve
satisfiability problems is the backtracking algorithm \cite{van2006backtracking}.
This technique has a very broad applicability and can be used whenever
it is possible to efficiently verify, using an oracle $P(x)$, if
a partial assignment $x$ is compatible with a solution or not. If
such $P(x)$ exists, the algorithm provides an efficient way to extend
partial configurations to valid assignments, with the effect of drastically
reducing the total number of the configurations to actually check.
In the backtracking technique, each variable can assume not only the
values $true/1$ or $false/0$, but also the value $indeterminate/*$:
If a variable is $indeterminate$, it does not participate in determining
the satisfiability of a clause. Thus, a clause is $indeterminate$
if it is not possible to decide with certainty if it is satisfied
or not. To make this point clearer, consider the following 1in3-SAT
clause $\mathcal{R}\left(x_{1},\,x_{2},\,x_{3}\right)$ that is satisfied
when exactly one variable among $x_{j}$ is $true$. In this case,
configurations like $\left\{ *,\,*,\,*\right\} $, $\left\{ 0,\,*,\,*\right\} $
or $\left\{ 1,\,0,\,*\right\} $ make the clause $\mathcal{R}$ $indeterminate$.
On the contrary, the clause $\mathcal{R}$ cannot be satisfied whenever
the configuration has already two ones, like for example $\left\{ 1,\,1,\,*\right\} $,
$\left\{ 1,\,*,\,1\right\} $ and $\left\{ *,\,1,\,1\right\} $. The
backtracking algorithm starts by setting the initial configuration
$x$ as completely $indeterminate$ $\left\{ *,\,*,\,\ldots,\,*\right\} $.
Then, following a heuristic $h(x)$ which gives the position $j$
of an $indeterminate$ variable, the configuration $x$ is expanded
by fixing $x_{j}$ to either $true$ or $false$. At this point, if
$P(x)$ returns that the partial configuration $x$ can lead to a
valid assignment, then another branching is made to the next $indeterminate$
variable. 
\begin{figure}
\begin{centering}
\includegraphics[width=0.95\textwidth]{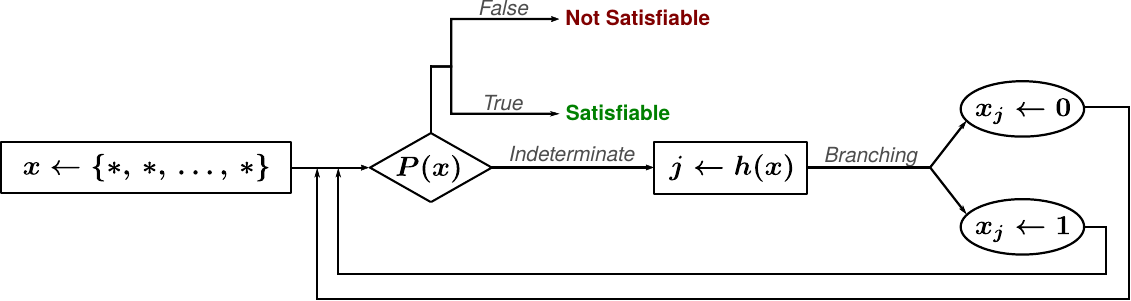}
\par\end{centering}

\caption{\textbf{\label{fig:backtracking}Graphical representation of the classical
backtracking algorithm.} The figure shows an example of backtracking
algorithm where each element of $x$ can assume three values: $\left\{ false/0,\,true/1,\,indeterminate/*\right\} $.
Here $P(x)$ is an oracle which returns \emph{false} if the assignment
x (either partial or not) violates any clause. Otherwise, $P(x)$
returns \emph{true} if $x$ is complete assignment and it satisfies
all the clauses, or \emph{indeterminate} if $x$ is partial and it
does not violate any clause. The oracle $h(x)$ return the position
of the next variable to explore in the decision tree given a partial
configuration $x$.}
\end{figure}
Otherwise, if an unsatisfied clause already exists, the branch is
``cut'' and the algorithm starts to explore another branch. The
set of partial configurations that the backtracking algorithms explores
is called ``decision tree''. The computational complexity of the
backtracking algorithm is therefore bounded by the size $T(n)$ of
the decision tree. In general, for satisfiability problems, $T(n)$
is expected to be exponential with the number of variables $n$. In
Figure~\ref{fig:backtracking}, a graphical illustration of the logic
behind the backtracking algorithm is depicted. 

Recently, an approach analogous to the backtracking algorithm has
been applied in the context of quantum computation \cite{montanaro2015quantum}.
The central idea of the quantum backtracking algorithm consists in
using a quantum walker \cite{childs2003exponential,ambainis2007quantum,szegedy2004quantum,magniez2011search}
to explore the tree of partial configurations and then ``mark''
a valid assignment for the satisfiability problem. More precisely,
given the two oracles $P(x)$ and $h(x)$ that can be evaluated in
$poly(n)$ operations, the quantum backtracking algorithm exhibits
a valid assignment in a number of oracle calls bounded by $O(\sqrt{T(n)})$,
which is quadratically faster than its classical counterpart. Observe
that the quantum backtracking algorithm provides the same quantum
speedup as the Grover algorithm. However, the quantum backtracking
algorithm can take advantage of the underlying structure of the satisfiability
problem, having an overall better performance since $T(n)\leq2^{n}$
and usually much smaller. In the rest of the Section, we show how
to apply the quantum backtracking algorithm to the proposed quantum
XOR-Reduction. We also compare the quantum XOR-Reduction algorithm
with the classical WalkSAT heuristic \cite{guidetti2011complexity}
and the quantum Adiabatic Quantum Optimization \cite{hen2011exponential}
to solve random instances of both 1in3-SAT and 2in4-SAT at the satisfiability
transition \cite{zdeborova2008constraint}, and provide numerical
evidence that our algorithm remains the fastest one.

As described in Section~\ref{sec:reduction-method}, the proposed
XOR-Reduction algorithm is based on a non-trivial (but polynomial
in both space and time) reduction so that any valid assignment $x$
of an Occupation problem is the combination (modulo 2) of $\left\{ \xi_{1},\,\xi_{2},\,\ldots,\,\xi_{k}\right\} $
linearly independent vectors plus the inhomogeneous solution $\bar{\xi}$
(see Eq.~(\ref{eq:XOR_space}) for more details), namely:
\begin{equation}
x\left(v_{1},\,\ldots,\,v_{k}\right)=v_{1}\xi_{1}\oplus v_{2}\xi_{2}\oplus\ldots\oplus v_{k}\xi_{k}\oplus\bar{\xi},\label{eq:XOR_space_2}
\end{equation}
with $\left\{ \xi_{1},\,\xi_{2},\,\ldots,\,\xi_{k}\right\} $ and
$\bar{\xi}$ depending on the specific instance $\mathcal{I}$ of
the Occupation problem. The size of the configuration space spanned
by $v$ is $2^{k}$ and, since $k=n-M^{\prime}$ (with $M^{\prime}$
representing the number of independent clauses), it results to be
effectively smaller than $2^{n}$. In Section~\ref{sec:worst-case}
we show that the worst-case complexity is bounded by $O(2^{k/2})$.
Nevertheless, we expect that the application of the quantum backtracking
algorithm to the space spanned by $\nu$ would lead, on average, to
a better performance. However, in order to apply the quantum backtracking
technique to the proposed XOR-Reduction algorithm, it is necessary
to rewrite the assignment $x$ in Eq.~(\ref{eq:XOR_space_2}) in
a form suitable to verify whether a partial assignment can be extended
to a solution or not. To achieve this goal, let us observe that one
can always construct (in polynomial time and space) a linear transformation
$U$ acting on the vector $v$ such that Eq.~(\ref{eq:XOR_space_2})
can be rewritten as
\begin{equation}
x\left(v_{1},\,\ldots,\,v_{k}\right)=KU^{-1}Uv\oplus\bar{\xi}=P\left(\begin{array}{c}
\mathbb{1}\\
H
\end{array}\right)Uv\oplus\bar{\xi}=P\left(\begin{array}{c}
v^{\prime}\\
Hv^{\prime}\oplus\bar{\xi}^{\prime}
\end{array}\right),\label{eq:XOR_space_3}
\end{equation}
with $P$ an appropriate permutation matrix that reorders the rows
of $K$. More precisely, the application of $U$ reduces the matrix
$K$ to a standard form comprising two blocks: The first $k$ rows
constitute the $k\times k$ identity matrix $\mathbb{1}$ while the
lower block corresponds to the $(n-k)\times k$ matrix $H$. Since
$v$ is arbitrary, we can directly express Eq.~(\ref{eq:XOR_space_3})
in terms of $v^{\prime}$ as
\begin{equation}
x_{P}\left(v_{1}^{\prime},\,\ldots,\,v_{k}^{\prime}\right)=\left(\begin{array}{c}
v^{\prime}\\
Hv^{\prime}\oplus\bar{\xi}^{\prime}
\end{array}\right),\label{eq:XOR_space_4}
\end{equation}
where $x_{P}$ and $x$ differ only by the permutation $P$, namely
only by a reorder of the variables. It is important to observe that,
after the transformation given by $U$, the assignment $x_{P}$ of
the Occupation problem is divided in two parts: The first $k$ variables
correspond exactly to the arbitrary reduced configuration $v^{\prime}$,
while the last $\left(n-k\right)$ variables are a linear combination
of the reduced configuration. The proposed quantum algorithm remains
exact even if it is combined with the quantum backtracking technique
\cite{montanaro2015quantum}. In addition, the use of the quantum
backtracking technique may give a better scaling on average than for
the worst-case: Indeed, if no bounds on $T(n)$ are known, the computational
complexity in the worst-case remains the same obtained by using the
Grover algorithm.

\begin{figure}
\begin{centering}
\includegraphics[width=0.75\textwidth]{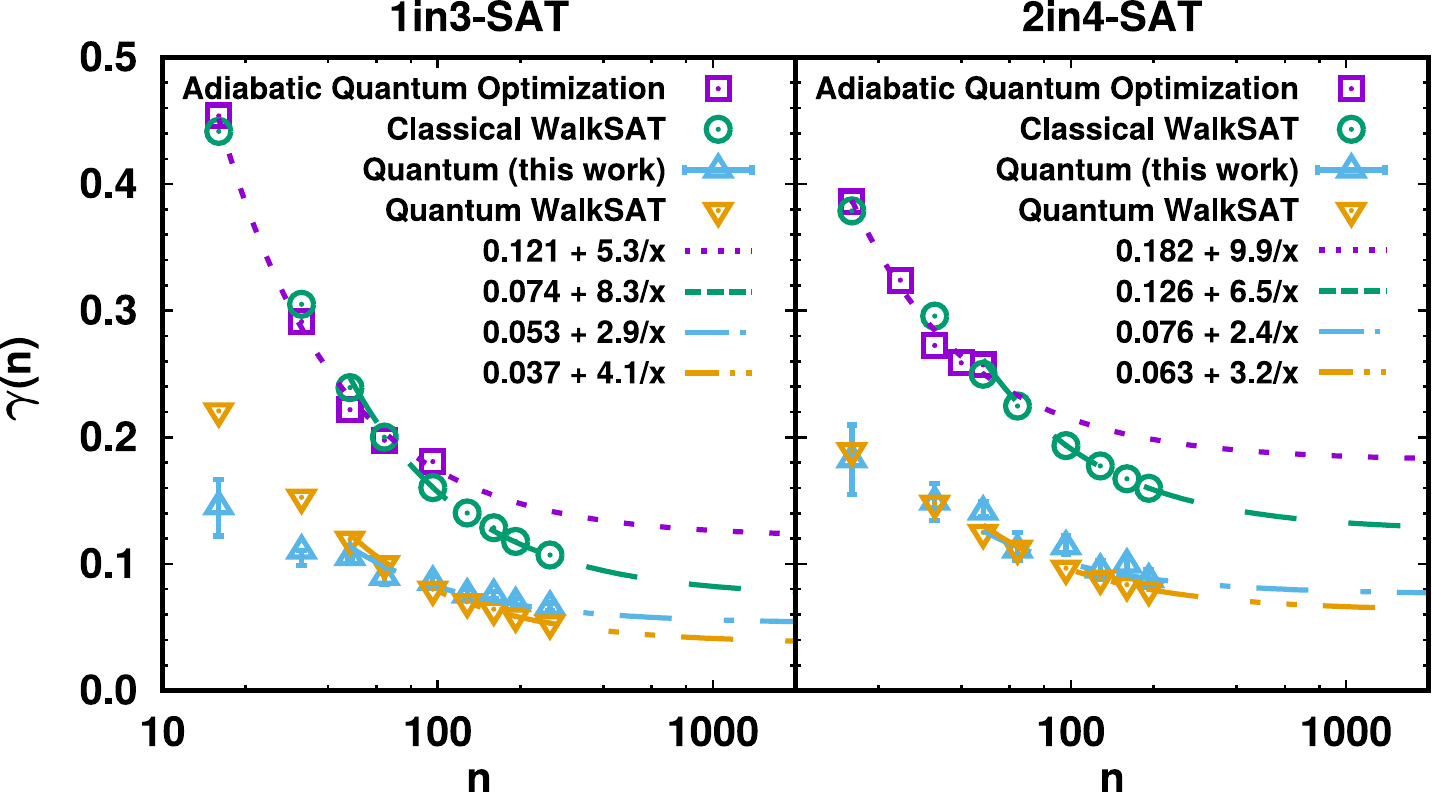}
\par\end{centering}

\caption{\textbf{\label{fig:scaling_typical_case}
The proposed quantum XOR-Reduction
algorithm is faster than both the classical WalkSAT heuristic and
the Adiabatic Quantum Optimization (AQO) at the satisfiability threshold.}
Comparison of the proposed quantum algorithm to both the well known
WalkSAT heuristic and Adiabatic Quantum Optimization (AQO), for locked
1in3-SAT and 2in4-SAT random instances at the satisfiability threshold
(which are respectively $\alpha_{SAT}=0.789$ and $\alpha_{SAT}=0.707$).
Numerical results for the classical WalkSAT and AQO have been extracted from
Ref. \cite{guidetti2011complexity} and Ref. \cite{hen2011exponential},
respectively. Results for the quantum WalkSAT have been obtained by assuming
that a quantum amplitude amplification \cite{ambainis2004quantum} 
would give a quadratic speedup with respect to
the classical results. 
Here, $\gamma(n)$ is either the computational time scaling (for WalkSAT
and AQO) or $\gamma(n)=\frac{1}{n}\log_{2}{\left\langle \sqrt{T(n)}\right\rangle }$,
with $T(n)$ the size of the decision tree at fixed number of variables
$n$. The average $\left\langle \cdot\right\rangle $ is computed
by sampling $1000$ instances for each $n$. The scaling has been
computed by fitting the numerical data (dashed lines intersects the
data points actually used in the fit). Error bars for the proposed
model represent the $10\%-90\%$ interval of confidence. As one can see,
the quantum version of the WalkSAT has a slightly better performance than
the proposed quantum algorithm. However, numerical results in 
\cite{hen2011exponential,guidetti2011complexity} have been obtained by
pre-selecting instances with at least a valid assignment. Therefore,
a worse performance is expected for both the classical and quantum 
WalKSAT heuristics when instances with no valid assignments are taken
into account.}
\end{figure}
In Figure~\ref{fig:scaling_typical_case}, we show the scaling of
the proposed quantum XOR-Reduction, when the quantum backtracking
technique is used, for random instances of locked 1in3-SAT and 2in4-SAT
at the satisfiability threshold (which are respectively $\alpha_{SAT}=0.789$
and $\alpha_{SAT}=0.707$), where typical instances are the hardest
to solve. 
In the implementation of the backtracking procedure, we have not optimized
over the order of exploration of the decision tree. 
Therefore, a better performance might be reached by considering
heuristics able to exploit the tree structure. In the figure, 
we also compare the quantum XOR-Reduction to the numerical 
results for the classical WalkSAT heuristic (numerical data have been extracted from 
Ref. \cite{guidetti2011complexity}) and for the Adiabatic Quantum Optimization 
(AQO) (numerical data have been extracted from Ref. \cite{hen2011exponential}). 
Unlike in \cite{hen2011exponential,guidetti2011complexity},
random instances are not pre-selected to have a unique solution and
may or may not have any solution. In the figure, $\gamma(n)$ is either
the computational time scaling (for WalkSAT and AQO) or 
$\gamma(n)=\frac{1}{n}\log_{2}{\left\langle \sqrt{T(n)}\right\rangle }$.
The mean value of $\left\langle \sqrt{T(n)}\right\rangle $
has been computed by averaging over $1000$ random instances. For each random
instance, the permutation matrix $P$ in Eq.~(\ref{eq:XOR_space_3})
has also been optimized in order to maximize the number of clauses
in which the $v^{\prime}$ variables appears, by running $100$ time
an optimization heuristic. The aforementioned reduction in 
Eq.~(\ref{eq:XOR_space_4}) can be done in polynomial time and space 
before exploring the decision tree.
As one can see, for both the satisfiability problems the quantum XOR-Reduction 
is the fastest among classical WalkSAT and AQO. 
In principle, the quantum amplitude amplification can be applied to 
the classical WalkSAT as well \cite{ambainis2004quantum}, obtaining a quadratic speedup with respect
to the classical performance. To have an idea of the performance of the 
quantum WalkSAT (without explicitly running the quantum algorithm), 
we simply divide the computational time scaling $\gamma(n)$
of the classical WalkSAT by a factor $2$. In this case, 
as shown in Figure~\ref{fig:scaling_typical_case}, the proposed quantum algorithm performs slightly worse.
However, it is important to stress that the WalkSAT is a widely used and hence well 
optimized algorithm while, for the proposed quantum algorithm, there is still space 
for improvement (for example, by proposing a better heuristic for exploring the decision tree). 
Moreover, it is important to mention that, whereas both WalkSAT and AQO are \emph{not} exact
and may potentially run forever if the instance does not admit any
valid assignment, the proposed quantum XOR-SAT reduction either provides
a solution or certifies that no solutions exist in the given bound.
In addition, the scaling proposed in Ref. \cite{guidetti2011complexity} is obtained by 
pre-selecting instances with at least one ground states. A worse 
scaling is expected when instances are randomly chosen with an $\alpha$ close to the 
transition threshold, value for which many instances do not have any valid assignment.
Finally, we want to stress that the proposed XOR-Reduction explores the XSAT configuration
space in a non-local fashion. Hence, it can 
potentially exploit long-range structures that are, instead, precluded to local 
search algorithms like WalkSAT and AQO.

\section{Conclusions}

In this work we have presented an exact quantum algorithm to solve
instances of Exact Satisfiability or, more generally, of Occupation
problems. The proposed quantum algorithm is based on a novel approach
which consists to identify a restricted subspace in which all the
valid assignments of an Occupation problem are contained (in our case,
the solution space of an appropriate XORSAT problem) and whose elements
can be efficiently enumerated. This approach led us to the development
of an algorithm able to solve a great variety of different Occupation
problems, as opposed to dedicated solvers that address specific problems.
Moreover, it can be potentially used to reduce the computational cost
of other satisfiability problems.

Regarding the worst-case scenario, we show that the proposed quantum
algorithm finds a valid assignment (or certify that none exists) using
only polynomial space and a number of oracle calls which is bounded
by $O(\sqrt{2^{n-M^{\prime}}})$, where $M^{\prime}\leq\min\left\{ n,\,M\right\} $
is the number of independent clauses. The proposed quantum algorithm
can also be extended to count the total number of valid assignments.
In this case, despite it still requires a polynomial number of resources, the
number of calls is bounded by $O(2^{n-M^{\prime}})$ instead. We compare
the worst-case scenario to the (known) best classical algorithms for
XSAT/1in3-SAT problems. Remarkably, we show that the proposed quantum
algorithm is the fastest one for sufficiently dense formula. It is
also interesting to observe that the proposed quantum algorithm monotonically
improves the performance by increasing the density of clauses. However,
this is not in contradiction with the na\"ive idea that instances
of satisfiability problems are (typically) easy to solve for either
very low or very dense formulas and hard to solve in the overlapping
region, since the worst-case bound is strictly algorithmic dependent
rather than problem dependent. As a concrete example, we provide the
worst-case bound to solve the Hamiltonian cycle by directly applying
our algorithm. More precisely, the quantum algorithm proposed in this
work can find a Hamiltonian cycle (or certify that none exists) for
3-regular graphs in a time bounded by $O\left({2^{n/4}}\right)$,
where $n$ is the number of nodes in the graph. The same worst-case
bound holds for $\left(3,3\right)$-regular bipartite graphs.

In addition, we showed that our quantum algorithm can be modified
to include techniques as the quantum backtracking
algorithm. We verified with numerical simulations that, on the typical
instance, the performance is indeed better than the expected worst-case
bound. Noteworthy, the proposed quantum algorithm remains the fastest
solver (compared to the classical WalkSAT heuristic and Adiabatic
Quantum Optimization) close to the satisfiability transition for the
locked 1in3-SAT and 2in4-SAT random instances.

\section{Acknowledgment}

The authors thank Eleanor G. Rieffel, Sergio Boixo, Ryan Babbush and
Itay Hen for the critical reading of the manuscript and several helpful
discussions. S.M and A.A-G were supported by NASA (Sponsor Award Number:
NNX14AF62G). G.G.G. and A.A-G acknowledge support from the Air Force
Office of Scientific Research through the grant 10323836-SUB.

\bibliographystyle{apsrevtitle}
\bibliography{quantum_exact_XSAT}

\end{document}